\documentclass[12pt]{article}
\usepackage{amsfonts}

\usepackage{amsmath}
\usepackage{amscd,amsmath,amssymb}

\newcommand{\be}{\begin{equation}}
\newcommand{\ee}{\end{equation}}
\newcommand{\bea}{\begin{eqnarray}}
\newcommand{\eea}{\end{eqnarray}}

\begin{document}

\thispagestyle{empty}
\vspace{3cm}
\begin{flushright}
%\texttt{hep-th/} \\[5mm]
\end{flushright}
\vspace{3cm}
\begin{center}
{\Large \textbf{A new $\mathcal{N}{=}8$ nonlinear supermultiplet }}
\end{center}

\vspace{1cm}

\begin{center}
{\large \textbf{S. Bellucci${}^{a}$, S. Krivonos${}^{b}$, A. Marrani${}^{c,
a}$ }}

${}^a$ \textit{INFN-Laboratori Nazionali di Frascati,Via Enrico Fermi 40,
00044 Frascati, Italy}\\[0pt]
\vspace{0.2cm}

\texttt{bellucci,marrani@lnf.infn.it} \vspace{0.4cm}

${}^b$ \textit{Bogoliubov Laboratory of Theoretical Physics, JINR, 141980
Dubna, Russia}\\[0pt]
\vspace{0.2cm}

\texttt{krivonos@theor\textbf{.}jinr.ru} \vspace{0.4cm}

${}^c$ \textit{Museo Storico della Fisica e Centro Studi e Ricerche ``Enrico
Fermi"\\[0pt]
Via Panisperna 89A, Compendio Viminale, 00184 Roma, Italy}
\end{center}

\vspace{2cm}
\begin{abstract}
We construct a new off-shell $\mathcal{N}{=}8$, $d{=}1$ nonlinear
supermultiplet $(\mathbf{4,8,4})$ proceeding from the nonlinear
realization of the $\mathcal{N}{=}8$, $d{=}1$ superconformal group
$OSp(4^{\star }|4)$ in its supercoset $\frac{OSp(4^{\star
}|4)}{SU(2)_{\mathcal{R}}\otimes \left\{ D,K\right\} \otimes
SO(4)}$. The irreducibility constraints for the superfields
automatically follow from appropriate covariant conditions on the
$osp(4^{\star }|4)$-valued Cartan superforms. We present the most
general sigma-model type action for $(\mathbf{4,8,4})$
supermultiplet. The relations between linear and nonlinear
$(\mathbf{4,8,4})$ supermultiplets and linear $\mathcal{N}{=}8$
$(\mathbf{5,8,3})$ vector supermultiplet are discussed.
\end{abstract}

\newpage \setcounter{page}{1}

\section{Introduction}

During the last few years it has become clear that in Supersymmetric Quantum
Mechanics (SQM) with extended $\mathcal{N}=4,8$ supersymmetries the
nonlinear supermultiplets play an essential role \cite{ikl1}-\cite{bks}. The
main reason for resort to nonlinearities is the presence of too strong restrictions on
the bosonic target-space metrics, in the case of theories with linear
supermultiplets. Indeed, the general consideration of $d=1$ sigma models
with $\mathcal{N}=4$ and $\mathcal{N}=8$ supersymmetries reveal the
following possible bosonic target geometries: hyper-K\"{a}hler with torsion,
for $\mathcal{N}=4$ supersymmetric theories with four physical bosons, and
octonionic-K\"{a}hler with torsion for $\mathcal{N}=8$ ones, in the case of
eight physical bosonic fields \cite{GPS}. Moreover, the detailed analysis of
the components and superfield actions for $\mathcal{N}=4,8$ cases with
diverse numbers of physical bosonic degrees of freedom shows that only
conformally flat geometries, with the additional restriction on the metrics of
bosonic manifolds to be harmonic functions, may arise \cite{N48}-\cite
{N48d}. Being quite general, these results keep open only a unique way to have
more complicated bosonic target-space geometries --- i.e. to introduce
nonlinear supermultiplets.

When dealing with nonlinear supermultiplets one should be able to overcome
at least two obstacles:

\begin{itemize}
\item  it is not clear how to find the proper superfield constraints
defining the irreducible nonlinear supermultiplets

\item  the construction of the invariant superfield actions is not evident.
\end{itemize}

One should mention that the dimensional reduction is not too useful for
obtaining the invariant $d=1$ superfield actions and the irreducible
constraints on the superfields. Although any $d{=}1$ super Poincar\'{e}
algebra can be obtained from a higher-dimensional one via dimensional
reduction, this is generally not true for $d{=}1$ super\textit{conformal\/}
algebras \cite{VP,FRS} and off-shell $d{=}1$ multiplets. For instance, no $d{%
=}4$ analog exists for the $\mathcal{N}{=}4$, $d{=}1$ multiplet with
off-shell content $(\mathbf{1,4,3})$ \cite{leva2} or $(\mathbf{3,4,1})$ \cite
{ismi,bepa}. Moreover, there exist off-shell $d{=}1$ supermultiplets
containing no auxiliary fields at all, something impossible for $d{\geq }3$
supersymmetry.

However, a convenient superfield approach to $d{=}1$ models which does not
resort to dimensional reduction and is self-contained in $d{=}1$ exists. It
is based on superfield nonlinear realizations of $d{=}1$ superconformal
groups. It was pioneered in \cite{leva2} and recently advanced in \cite
{ikl1,ikl2,bikl1}. In this approach the physical bosons and fermions,
together with the $d{=}1$ superspace coordinates, prove to be coset
parameters associated with the appropriate generators of the superconformal
group. The conditions which identify the fermionic components of the bosonic
superfields with the cosets fermionic parameters are just the irreducible
constraints singling out the proper supermultiplets.

Using the nonlinear realizations approach, in \cite{ikl1} all known
linear off-shell multiplets of $\mathcal{N}=4$, $d{=}1$ Poincar\'{e}
supersymmetry were recovered and a two novel nonlinear ones were
found. Concerning $\mathcal{N}{=}8,d{=}1$ supermultiplets, in
\cite{bikl1} a similar analysis has been started along the same
line. It has been shown that the $(\mathbf{5,8,3})$ and
$(\mathbf{3,8,5})$ multiplets come out as
the Goldstone ones, parameterizing the specific cosets of the supergroup $%
OSp(4^{\star }|4)$. Consequently, in \cite{bikl2} a superfield description
of all other linear off-shell $\mathcal{N}{=}8$, $d{=}1$ supermultiplets
with $\mathbf{8}$ fermions, in both $\mathcal{N}{=}8$ and $\mathcal{N}{=}4$
superspaces, was given. Finally, the first $\mathcal{N}=8$ nonlinear
supermultiplet $(\mathbf{2,8,6})$ has been constructed in \cite{bbks}.
However, the task of deriving an exhaustive list of off-shell $\mathcal{N}{=}%
8$ supermultiplets and the relevant constrained $\mathcal{N}{=}8$, $d{=}1$
superfields is much more complicated as compared to the $\mathcal{N}{=}4$
case, in view of the existence of many non-equivalent $\mathcal{N}{=}8$
superconformal groups ($OSp(4^{\star }|4)$, $OSp(8|2)$, $F(4)$ and $SU(1,1|4)
$, see e.g. \cite{VP}), with a large number of different coset
supermanifolds. Moreover, the explicit construction of linear \cite{286} and
nonlinear \cite{bbks} $(\mathbf{2,8,6})$ supermultiplets demonstrates that
the constraints which follow from the nonlinear realization approach should be
accompanied by additional, second order in spinor covariant derivatives,
constraints, in order to define irreducible $\mathcal{N}=8$ supermultiplets.

The subject of the present paper is the study of the nonlinear $\mathcal{N}{=%
}8$ supermultiplet with field content $(\mathbf{4,8,4})$ obtained by a
nonlinear realization of the $OSp(4^{\star }|4)$ supergroup on a particular
coset supermanifold $\frac{OSp(4^{\star }|4)}{SU(2)_{\mathcal{R}}\otimes
\left\{ D,K\right\} \otimes SO(4)}$. After reviewing some basic facts on the
$\mathcal{N}{=}8$, $d{=}1$ superspace and the Lie superalgebra $osp(4^{\star
}|4)$ in Section 2 we give a $\mathcal{N}{=}8$ superfield formulation of
the nonlinear multiplet $(\mathbf{4,8,4})$. In Section 3 we obtain the
superfield constraints defining irreducible linear and nonlinear $(\mathbf{%
4,8,4})$ supermultiplets by the reduction procedures from the $(\mathbf{5,8,3%
})$ vector multiplet. Using these results in Section 4 we construct the
most general action for our nonlinear supermultiplet and provide a detailed
analysis of its bosonic part. A summary of our results and an outlook are
the contents of the concluding Section 5.

\setcounter{equation}0

\section{The $\mathcal{N}{=}8$ $\mathbf{(4,8,4)}$ nonlinear multiplet}

Like the $\mathcal{N}=8$ tensor and vector supermultiplets \cite{bikl1}, the
new nonlinear $\mathcal{N}=8$ multiplet we are going to consider can be
obtained from a proper nonlinear realization of the $\mathcal{N}=8$, $d=1$
superconformal group $OSp(4^{\star }|4)$ in $\mathcal{N}{=}8$, $d{=}1$
superspace. After exposing some basic facts in Subsections 2.1 and 2.2, in
Subsection 2.3 we will give the details of the relevant nonlinear
realization procedure giving rise to the nonlinear $\mathbf{(4,8,4)}$
multiplet\footnote{%
We use the notation \textbf{(m,8,8-m)} to identify an off-shell $\mathcal{N}{%
=}8$, $d{=}1$ supermultiplet with \textbf{m} physical bosons, \textbf{8}
fermions and \textbf{8-m} auxiliary bosonic components.}. Our basic
notations follow those of Ref. \cite{bikl1}.

\subsection{The $\mathcal{N}{=}8$, $d{=}1$ superspace}

The eight real Grassmann coordinates of $\mathcal{N}{=}8,d{=}1$ superspace $%
\mathbb{R}^{(1|8)}$ can be arranged into one of three 8-dimensional real
irreps of $SO(8)$ -- the maximal automorphism group of $\mathcal{N}{=}8,d{=}1
$ super Poincar\'{e} algebra. However, the constraints defining the
irreducible $\mathcal{N}{=}8$ supermultiplets in general break this $SO(8)$
symmetry. So, it is preferable to split the 8 coordinates into two real
quartets
\begin{equation}
\mathbb{R}^{(1|8)}=(t,\,\theta _{ia},\,\vartheta _{\alpha A})\,,\qquad \quad
\overline{\left( \theta _{ia}\right) }=\theta ^{ia},\;\overline{\left(
\vartheta _{\alpha A}\right) }=\vartheta ^{\alpha A},\quad i,a,\alpha
,A=1,2\,,  \label{def1}
\end{equation}
in terms of which only four commuting automorphism $SU(2)$ groups will be
explicit. The further symmetry breaking can be understood as the identification
of some of these $SU(2)$, whereas additional symmetries, if existing, mix different $%
SU(2)$ indices. The corresponding covariant derivatives are defined by
\begin{equation}
D^{ia}=\frac{\partial }{\partial \theta _{ia}}+i\theta ^{ia}\partial
_{t}\;,\;\nabla ^{\alpha A}=\frac{\partial }{\partial \vartheta _{\alpha A}}%
+i\vartheta ^{\alpha A}\partial _{t}\;.  \label{def2}
\end{equation}
By construction, they obey the algebra:\footnote{%
We use the following convention for the skew-symmetric tensor $\epsilon $: $%
\;\epsilon _{ij}\epsilon ^{jk}=\delta _{i}^{k}\;,\quad \epsilon
_{12}=\epsilon ^{21}\equiv 1\;$.}
\begin{equation}
\left\{ D^{ia},D^{jb}\right\} =2i\epsilon ^{ij}\epsilon ^{ab}\partial
_{t}\,,\quad \left\{ \nabla ^{\alpha A},\nabla ^{\beta B}\right\}
=2i\epsilon ^{\alpha \beta }\epsilon ^{AB}\partial _{t}\;.  \label{def3}
\end{equation}
Thus all our $\mathcal{N}=8,d=1$ superfields depend on $(t,\,\theta
_{ia},\,\vartheta _{\alpha A})$ and the differential constraints on the
relevant superfields will be defined directly in terms of the spinor
covariant derivatives (\ref{def2}).

\subsection{The superalgebra $osp(4^{\star }|4)$}

Let us briefly recall some basic facts about the Lie superalgebra $%
osp(4^{\star }|4)$ \cite{{FRS},{bikl1}}. It contains the following sixteen
spinor generators:
\begin{equation}
Q_{1}^{iaA},\;Q_{2}^{i\alpha A},\quad \overline{\left( Q^{iaA}\right) }%
=\epsilon _{ij}\epsilon _{ab}Q^{jbA},\quad (i,a,\alpha ,A=1,2),  \label{alg1}
\end{equation}
and sixteen bosonic generators:
\begin{equation}
T_{0}^{AB},\quad T^{ij},\quad T_{1}^{ab},\quad T_{2}^{\alpha \beta },\quad
U^{a\alpha }\;.  \label{alg2}
\end{equation}
Here, the indices $A,i,a$ and $\alpha $ refer to fundamental representations
of the mutually commuting $sl(2,\mathbb{R})\sim T_{0}^{AB}$ and three $%
su(2)\sim T^{ij},T_{1}^{ab},T_{2}^{\alpha \beta }$ algebras. The four
generators $U^{a\alpha }$ belong to the coset $SO(5)/SO(4)$ with $SO(4)$
generated by $T_{1}^{ab}$ and $T_{2}^{\alpha \beta }\,$.

The commutators of any $SU(2)$-generators with $Q$ have the standard form
\begin{equation}
\left[ T^{ab},Q^{c}\right] =-\frac{i}{2}\left( \epsilon ^{ac}Q^{b}+\epsilon
^{bc}Q^{a}\right) ,  \label{alg5}
\end{equation}
where $a,b$ refer to some particular sort of indices (with other indices
of $Q$ being suppressed).

The commutators with the coset $SO(5)/SO(4)$ generators $U^{a\alpha }$ mix
the $Q_{1}^{iaA}$ and $Q_{2}^{i\alpha A}$ generators
\begin{equation}
\left[ U^{a\alpha },Q_{1}^{ibA}\right] =-i\epsilon ^{ab}Q_{2}^{i\alpha
A},\quad \left[ U^{a\alpha },Q_{2}^{i\beta A}\right] =-i\epsilon ^{\alpha
\beta }Q_{1}^{iaA}\;.  \label{alg6}
\end{equation}
Finally, the anticommutators of the fermionic generators read~
\begin{eqnarray}
&&\left\{ Q_{1}^{iaA},Q_{1}^{jbB}\right\} =-2\left( \epsilon ^{ij}\epsilon
^{ab}T_{0}^{AB}-2\epsilon ^{ij}\epsilon ^{AB}T_{1}^{ab}+\epsilon
^{ab}\epsilon ^{AB}T^{ij}\right) ,  \notag  \label{alg7} \\[6pt]
&&\left\{ Q_{2}^{i\alpha A},Q_{2}^{j\beta B}\right\} =-2\left( \epsilon
^{ij}\epsilon ^{\alpha \beta }T_{0}^{AB}-2\epsilon ^{ij}\epsilon
^{AB}T_{2}^{\alpha \beta }+\epsilon ^{\alpha \beta }\epsilon
^{AB}T^{ij}\right) ,  \notag \\[6pt]
&&\left\{ Q_{1}^{iaA},Q_{2}^{j\alpha B}\right\} =2\epsilon ^{ij}\epsilon
^{AB}U^{a\alpha }\;.
\end{eqnarray}
For what follows it is convenient to pass to another notation,
\begin{eqnarray}
&&P\equiv T_{0}^{22},\;K\equiv T_{0}^{11},\;D\equiv -T_{0}^{12},\quad
V\equiv T^{22},\;{\overline{V}}\equiv T^{11},\;V_{3}\equiv T^{12},  \notag \\%
[6pt]
&&Q^{ia}\equiv -Q_{1}^{ia2},\;\mathcal{Q}^{i\alpha }\equiv -Q_{2}^{i\alpha
2},\;S^{ia}\equiv Q_{1}^{ia1},\;\mathcal{S}^{i\alpha }\equiv Q_{2}^{i\alpha
1}\,.  \label{alg8}
\end{eqnarray}
One can check that $P$ and $Q^{ia},\mathcal{Q}^{i\alpha }$ constitute a $%
\mathcal{N}=8,d=1$ Poincar\'{e} superalgebra. The generators $D,K$ and $%
S^{ia},\mathcal{S}^{i\alpha }$ stand for the $d=1$ dilatations, special
conformal transformations and conformal supersymmetry, respectively.

\subsection{A new supercoset of $OSp(4^{\star }|4)$}

Our goal is to construct the nonlinear supermultiplet with off-shell content
$(\mathbf{4,8,4})$. In the nonlinear realization approach the physical
bosonic components parameterize some coset of the given supergroup. So the
first task is to identify such a four-dimensional bosonic coset in the
supergroup $OSp(4^{\star }|4)$. One of the possible choices is the
supercoset $\frac{OSp(4^{\star }|4)}{U(1)_{\mathcal{R}}\otimes SO(5)}
$. For this case the bosonic Goldstone superfields parameterize the coset $%
D\otimes SU(2)$. The corresponding supermultiplet includes the dilaton and
three fields living on the sphere $SU(2)$. It is just the linear $(\mathbf{4,8,4%
})$ supermultiplet \cite{bikl2}.

Another possibility is to consider the supercoset $\frac{OSp(4^{\star }|4)}{%
SU(2)_{\mathcal{R}}\otimes \left\{ D,K\right\} \otimes SO(4)}$. As it can be
easily seen, it contains the bosonic coset $SO(5)/SO(4)$;
the four physical bosonic fields of the resulting multiplet are nothing but
the parameters describing the 4-sphere $S^{4}=SO(5)/SO(4)$. Differently from
the previously mentioned supercoset, the dilaton associated with the
dilatation generator $D$ will not appear. Therefore, despite the fact that
the superconformal group $OSp(4^{\star }|4)$ is perfectly realized on our
supercoset there is no possibility to construct superconformally invariant
action in our case, because without dilaton there is no possibility to
compensate the dilatonic weight of the superspace measure. Nevertheless, the
$\mathcal{N}=8$ supersymmetric sigma-model type of the action can be
constructed for the supermultiplet in question.

Thus, we are going to realize the superconformal group $OSp(4^{\star }|4)$
in the coset superspace $\frac{OSp(4^{\star }|4)}{SU(2)_{\mathcal{R}}\otimes
\left\{ D,K\right\} \otimes SO(4)}$ parameterized as
\begin{equation}
g=e^{itP}e^{\theta _{ia}Q^{ia}+\vartheta _{i\alpha }\mathcal{Q}^{i\alpha
}}e^{\psi _{ia}S^{ia}+\xi _{i\alpha }\mathcal{S}^{i\alpha }}e^{iv_{\alpha
a}U^{\alpha a}}~.  \label{ncoset}
\end{equation}
As usual, in order to find the covariant irreducibility conditions on the
coset superfields, we must impose the inverse Higgs constraints \cite{IH} on
the left-covariant $osp(4^{\star }|4)$-valued Cartan one-form $%
\Omega=g^{-1}dg$. Concerning the treated case, the relevant constraints are
\begin{equation}
\left. \omega _{U}^{\alpha a}\right| =0\;,  \label{nih}
\end{equation}
where $|$ denotes the spinor projection. These constraints are manifestly
covariant under the left action of the whole supergroup $OSp(4^{\star }|4)$.
Indeed, with respect to the action of this supergroup by a left
multiplications on the coset element (\ref{ncoset}), the Cartan forms are
rotated by the elements of the stability subgroup $SU(2)_{\mathcal{R}%
}\otimes \left\{ D,K\right\} \otimes SO(4)$. Clearly, the constraints (\ref
{nih}) are invariant under such rotations. Explicitly, the Cartan form $%
\omega _{U}^{\alpha a}$ reads
\begin{equation}
\omega _{U}^{\alpha a}=\frac{2}{2+V^{2}}\left[ idV_{\alpha a}+\left( \delta
_{a}^{b}\delta _{\alpha }^{\beta }+V_{a}^{\beta }V_{\alpha }^{b}\right)
\Omega _{\beta b}+V_{\alpha }^{c}\omega _{ca}+V_{a}^{\beta }\omega _{\beta
\alpha }\right] ,  \label{nform}
\end{equation}
where
\begin{equation}
\Omega _{\alpha a}=-2\left( d\vartheta _{i\alpha }\psi _{a}^{i}+d\theta
_{ia}\xi _{\alpha }^{i}\right) ,\;\omega _{ab}=-4d\theta _{i(a}\psi
_{b)}^{i},\;\omega _{\alpha \beta }=-4d\vartheta _{i(\alpha }\xi _{\beta
)}^{i}
\end{equation}
and
\begin{equation}
V_{\alpha a}=\frac{\tan \sqrt{\frac{v^{2}}{2}}}{\sqrt{\frac{v^{2}}{2}}}%
v_{\alpha a},\text{ \ \ }v^{2}=\epsilon ^{ab}\epsilon ^{\alpha \beta
}v_{\alpha a}v_{\beta b}.  \label{VV}
\end{equation}
Selecting the $d\theta_{ia}$ and $d\vartheta _{i\alpha }$ projections of the
constraints (\ref{nih}) we will get
\begin{eqnarray}
&&iD^{ib}V_{\alpha a}+2\delta _{a}^{b}\left( \xi _{\alpha }^{i}+V_{\alpha
}^{c}\psi _{c}^{i}\right) +2V_{\alpha }^{b}\left( \psi _{a}^{i}+V_{a}^{\beta
}\xi _{\beta }^{i}\right) =0,  \label{eqs1} \\
&&  \notag \\
&&i\nabla ^{i\beta }V_{\alpha a}+2V_{a}^{\beta }\left( \xi _{\alpha
}^{i}+V_{\alpha }^{c}\psi _{c}^{i}\right) +2\delta _{\alpha }^{\beta }\left(
\psi _{a}^{i}+V_{a}^{\gamma }\xi _{\gamma }^{i}\right) =0.  \label{eqs2}
\end{eqnarray}
Eqs. (\ref{eqs1})-(\ref{eqs2}) allow one to express the eight fermions $\psi
_{a}^{i},\xi _{\alpha }^{i}$ in terms of the covariant derivatives of the
four bosonic superfields $V_{\alpha a},$ and therefore such equations properly
constrain the $V_{\alpha a}$'s. These constraints may be written in two
equivalent form
\begin{equation}
\left( D^{i(a}-V_{\beta }^{(a}\nabla ^{i\beta }\right) V_{\alpha
}^{b)}=0,\quad \left( \nabla ^{i(\alpha }-V_{b}^{(\alpha }D^{ib}\right)
V_{a}^{\beta )}=0,\label{nconstr1}
\end{equation}
or
\begin{equation}
D^{i(a}X_{\alpha }^{b)}-X^{\beta (a}\nabla _{\alpha }^{i}X_{\beta
}^{b)}=0,\quad \nabla ^{i(\alpha }X_{a}^{\beta )}-X^{b(\alpha
}D_{a}^{i}X_{b}^{\beta )}=0,  \label{nconstr2}
\end{equation}
where, as usual, the round brackets denote the symmetrization of the enclosed
indices, and we introduced
\begin{equation}
X_{a\alpha }\equiv \frac{2}{2-V^{2}}V_{a\alpha }.
\end{equation}
As it can be seen, such constraints are nonlinear, and therefore the
considered $\mathcal{N}=8$, $d=1$ multiplet may be referred to as the $%
\mathbf{(4,8,4)}$ nonlinear supermultiplet. Let us observe that discarding of the
nonlinear terms in (\ref{nconstr2}) yields the linear $\mathbf{(4,8,4)}$
supermultiplet \cite{bikl2}.

Besides ensuring the covariance of the constraints (\ref{nconstr2}), the
coset approach gives the easiest way to find the transformation properties
of the coordinates and superfields under the supergroup $OSp(4^{\star }|4)$.

The $\mathcal{N}=8,d=1$ Poincar\`{e} supersymmetry is realized in the
standard way
\begin{equation}
\delta t=-i\left( \eta _{ia}\theta ^{ia}+\eta _{i\alpha }\vartheta ^{i\alpha
}\right) ,\quad \delta \theta _{ia}=\eta _{ia},\;\delta \vartheta _{i\alpha
}=\eta _{i\alpha }  \label{n8p}
\end{equation}
and $V_{a\alpha }$ is a scalar with respect to these transformations. For
what concerns the transformation properties of the coordinates and
superfields $V^{\alpha a}$ under the conformal supersymmetry generated by
the left action of the element
\begin{equation}
g_{1}=e^{\eta _{ia}S^{ia}+\eta_{i\alpha }\mathcal{S}^{i\alpha }},
\end{equation}
one should note that the coordinates of the superspace are transformed in the
same way as in \cite{bikl1}
\begin{eqnarray}
&&\delta t=-it\left( \eta ^{ia}\theta _{ia}+\eta ^{i\alpha }\vartheta
_{i\alpha }\right) +\left( \eta _{a}^{i}\theta ^{ja}+\eta _{\alpha
}^{i}\vartheta ^{j\alpha }\right) \left( \theta _{ib}\theta
_{j}^{b}+\vartheta _{i\beta }\vartheta _{j}^{\beta }\right) \;,  \notag
\label{n8s} \\[6pt]
&&\delta \theta _{ia}=t\eta _{ia}-i\eta _{a}^{j}\theta _{jb}\theta
_{i}^{b}+2i\eta _{b}^{j}\theta _{i}^{b}\theta _{ja}-i\eta _{a}^{j}\vartheta
_{j\alpha }\vartheta _{i}^{\alpha }+2i\eta _{j}^{\alpha }\vartheta _{i\alpha
}\theta _{a}^{j}\;,  \notag \\[6pt]
&&\delta \vartheta _{i\alpha }=t\eta _{i\alpha }-i\eta _{\alpha
}^{j}\vartheta _{j\beta }\vartheta _{i}^{\beta }+2i\eta _{\beta
}^{j}\vartheta _{i}^{\beta }\vartheta _{j\alpha }-i\eta _{\alpha }^{j}\theta
_{ja}\theta _{i}^{a}+2i\eta _{a}^{j}\theta _{i}^{a}\vartheta _{j\alpha }\;,
\end{eqnarray}
while the superfield $V_{a\alpha }$ transform as
\begin{equation}
\delta V_{\alpha a}=2i\left( \delta _{\alpha }^{\beta }\delta
_{a}^{b}+V_{\alpha }^{b}V_{a}^{\beta }\right) A_{\beta b}+2iA_{ab}V_{\alpha
}^{b}+2iA_{\alpha \beta }V_{a}^{\beta }\,,
\end{equation}
where
\begin{equation}
A_{\alpha a}=\theta _{ia}\eta_{\alpha }^{i}+\vartheta _{i\alpha
}\eta _{a}^{i}\;,\;A_{ab}=\theta _{ia}\eta _{b}^{i}+\theta _{ib}\eta
_{a}^{i}\;,\;A_{\alpha \beta }=\vartheta _{i\alpha }\eta_{\beta
}^{i}+\vartheta _{i\beta }\eta_{\alpha }^{i}\;.
\end{equation}
The transformations with respect to other generators of the supergroup $%
OSp(4^{\star }|4)$ can be easily found from (\ref{n8p}), (\ref{n8s}) since
all bosonic transformations appear in the anticommutators of the conformal
and Poincar\'{e} supersymmetries. For the reader's convenience we will present
here the explicit form of the transformations under the left action of the $%
SO(5)/SO(4)$ element represented by
\begin{equation}
g_{2}=e^{ia_{\alpha a}U^{\alpha a}}\,
\end{equation}
which read as follows:
\begin{equation}
\delta \theta _{ia}=a_{\alpha a}\vartheta _{i}^{\alpha }\;,\quad \delta
\vartheta _{i\alpha }=a_{\alpha a}\theta _{i}^{a}\;;
\end{equation}
\begin{equation}
\delta V_{a\alpha }=\left( 1-\frac{V^{2}}{2}\right) a_{\alpha a}+a_{\beta
b}V^{b\beta }V_{a\alpha },\text{ \ \ \ }\delta X_{a\alpha }=a_{\alpha
a}+2a_{\beta b}X^{b\beta }X_{a\alpha }.
\end{equation}

Thus, the quartet of the $\mathcal{N}=8$ bosonic superfields $V_{ia}$
subjected to the nonlinear constraints (\ref{nconstr1}) defines the nonlinear
$\mathbf{(4,8,4)}$ supermultiplet.

\setcounter{equation}0

\section{Reduction from $\mathcal{N}=8$ vector supermultiplet and $\mathcal{N%
}=4$ superfield formulation}

Our construction of the nonlinear $\mathbf{(4,8,4)}$ supermultiplet is
very similar to the consideration of the $\mathbf{(5,8,3)}$ supermultiplet in
\cite{bikl1}. The only, though crucial difference, is the absence of the
dilaton among the components of our superfields $V_{ia}$. One may wonder
whether it is possible to reconstruct the nonlinear $\mathbf{(4,8,4)}$
supermultiplet by a direct reduction from the $\mathbf{(5,8,3)}$ one, as in the
case of $\mathcal{N}=4$ supermultiplets \cite{root}. Next we
demonstrate that such reduction indeed exists. Moreover, there are two
different reductions from $\mathcal{N}=8$, $d=1$ vector multiplet $\mathbf{%
(5,8,3)}$ to the supermultiplets $\mathbf{(4,8,4)}$ -- one reproducing the
linear supermultiplets, while second giving rise to the nonlinear one.

\subsection{Two reductions from $\mathcal{N}=8$ vector supermultiplet}

In order to properly analyze such reductions, it is convenient to recall
some basic facts on the $\mathcal{N}=8$ vector multiplet (see \cite{bikl1}
for further elucidation). The $\mathcal{N}=8$ multiplet $(\mathbf{5,8,3})$,
already considered in \cite{DE}, has been obtained in \cite{bikl1} from a
nonlinear realization of the same $\mathcal{N}=8$, $d=1$ superconformal
group $OSp(4^{\star }|4)$ in the coset superspace $\frac{OSp(4^{\star }|4)}{%
SU(2)_{\mathcal{R}}\otimes SO(4)}$ parameterized as
\begin{equation}
g=e^{itP}e^{\theta _{ia}Q^{ia}+\vartheta _{i\alpha }\mathcal{Q}^{i\alpha
}}e^{\psi _{ia}S^{ia}+\xi _{i\alpha }\mathcal{S}^{i\alpha
}}e^{izK}e^{iuD}e^{iv_{\alpha a}U^{\alpha a}}~.  \label{5coset}
\end{equation}
Beside the 4-dimensional bosonic coset $SO(5)/SO(4)$,
the physical bosonic field content of the vector multiplet includes the
dilaton superfield associated with the generator $D$. In this case, the
invariant constraints read
\begin{equation}
\omega _{D}=0\;,\quad \left. \omega _{U}^{\alpha a}\right| =0\;.  \label{5ih}
\end{equation}
Thus we see that, besides the same constraints on the Cartan forms $%
SO(5)/SO(4)$, there is an additional one which nullifies the dilaton form $\omega
_{D}$. The constraints (\ref{5ih}) allow one to express the Goldstone spinor
superfields and the boost superfield $z$ in terms of the spinor and $t$%
-derivatives of the remaining bosonic Goldstone superfields $u,v_{\alpha a}$%
. Moreover, they also imply the following irreducibility constraints:
\begin{equation}
D^{ib}\mathcal{V}_{\alpha a}+\delta _{a}^{b}\nabla _{\alpha }^{i}\mathcal{U}%
=0\;,\quad \nabla ^{i\beta }\mathcal{V}_{\alpha a}+\delta _{\alpha }^{\beta
}D_{a}^{i}\mathcal{U}=0\;,  \label{5constr}
\end{equation}
where
\begin{equation}
\mathcal{V}_{\alpha a}=e^{-u}\frac{2V_{\alpha a}}{2+V^{2}}\,,\quad \mathcal{U%
}=e^{-u}\left( \frac{2-V^{2}}{2+V^{2}}\right) \,,  \label{5Lambda}
\end{equation}
with $V_{\alpha a}$ defined in (\ref{VV}).

Let us now consider the reductions of the $\mathbf{(5,8,3)}$ vector
multiplet.

The first reduction procedure is rather trivial. We can start by replacing $%
D_{a}^{i}\mathcal{U}$ and $\nabla _{\alpha }^{i}\mathcal{U}$ in (\ref{5ih})
by arbitrary fermionic superfields $\Psi _{i}^{a}$ and $\Xi _{\alpha }^{i}$;
Eqs. (\ref{5ih}) will thence define such superfields in terms of covariant
spinor derivatives of the Goldstone bosonic superfields $\mathcal{V}_{\alpha
a}$, by constraining them as follows:
\begin{equation}
D^{i(a}\mathcal{V}_{\alpha }^{b)}=0,\quad \nabla ^{i(\alpha }\mathcal{V}%
_{a}^{\beta )}=0\;.
\end{equation}
Such constraining conditions are nothing but the ones defining the $%
\mathcal{N}=8$ linear $\mathbf{(4,8,4)}$ supermultiplet \cite{bikl2}. From
the previously performed replacement, it is clear that this reduction
procedure corresponds to \textit{``}removing'' the first bosonic component
of the superfield $\mathcal{U}$ from the set of physical bosons and
replacing it by an auxiliary field.

The second reduction corresponds to the \textit{``}removal'' of the real
dilaton superfield $u$. It is clear from (\ref{5Lambda}) that, in order to do
this, one has to define the new superfields $X_{\alpha a}$ as follows:
\begin{equation}
X_{\alpha a}\equiv \frac{\mathcal{V}_{\alpha a}}{\mathcal{U}}.  \label{X}
\end{equation}
The rewriting of the constraints (\ref{5ih}) in terms of $X_{\alpha a}$
gives rise to nothing else than the constraints (\ref{nconstr2}).

Summarizing, starting from the multiplet $\mathbf{(5,8,3)}$, the dimensional
reduction along the first bosonic component of the superfield $\mathcal{U}$
yields the linear $\mathbf{(4,8,4)}$ supermultiplet \cite{bikl2}, whereas
the removal of the dilaton $u$ yields the previously introduced nonlinear $%
\mathbf{(4,8,4)}$ multiplet. The existence of such a reduction is very
useful for the construction of the superfield action (see Section 4). Here
we will use this reduction, in order to provide a $\mathcal{N}=4$ description of our
nonlinear supermultiplet.

\subsection{$\mathcal{N}=4$ superfield formulations}

The use of the $\mathcal{N}=8$ superfield formalism is rather convenient when
considering the transformation properties, the invariance of the basic
constraints, etc. At the same time the $\mathcal{N}=4$ superspace
description is preferable for constructing the action. In order to find the $%
\mathcal{N}=4$ superfields content of our nonlinear supermultiplet we will
use its previously established connection with the linear $\mathbf{(5,8,3)}$
supermultiplet.

In order to formulate the nonlinear $\mathbf{(4,8,4)}$ supermultiplet in
terms of $\mathcal{N}=4$ superfields, it is convenient to recall the $%
\mathcal{N}=4$ splitting of the $\mathcal{N}=8$ vector multiplet \cite{bikl2}%
. For our purposes, we just need to define all superfields in the $\mathcal{N}%
=4$, $d=1$ superspace $\mathbb{R}^{(1\mid 4)}$ which is parameterized by the
coordinates $\left\{ t,\theta _{ia}\right\} $. The constraints (\ref{5ih})
imply that the spinor derivatives of all involved superfields with respect
to $\vartheta _{i\alpha }$ are expressed in terms of the spinor derivatives
with respect to $\theta _{ia}$. Consequently, the essential $\mathcal{N}=4$
superfield components in the $\vartheta $-expansion of the physical
Goldstone bosonic superfields $\mathcal{V}_{\alpha a}$ and $\mathcal{U}$ of
the vector multiplet are only the first ones
\begin{equation}
\hat{\mathcal{V}}_{\alpha a}\equiv \mathcal{V}_{\alpha a}|_{\vartheta
=0}\;,\quad \hat{\mathcal{U}}\equiv \mathcal{U}|_{\vartheta =0}\,.
\label{n4comp}
\end{equation}
These five bosonic $\mathcal{N}=4$ superfields, expressing the whole
off-shell component content of the $(\mathbf{5,8,3})$ vector multiplet, are
subjected by (\ref{5ih}) to the following irreducibility constraints in $%
\mathbb{R}^{(1\mid 4)}$ \cite{bikl2}:
\begin{equation}
D^{i(a}\hat{\mathcal{V}}{}^{b)\alpha }=0,\quad D^{i(a}D_{i}^{b)}\hat{%
\mathcal{U}}=0.  \label{5constra}
\end{equation}
Thus, by adopting such a $\mathcal{N}=4$ superspace perspective, the $%
\mathcal{N}=8$ vector supermultiplet may be considered as the sum of the $%
\mathcal{N}=4$, $d=1$ hypermultiplet $\hat{\mathcal{V}}_{\alpha a}$ (with $(%
\mathbf{4,4,0})$ off-shell component content) and the $\mathcal{N}=4$
``old'' tensor multiplet $\hat{\mathcal{U}}$ (with $(\mathbf{1,4,3})$
content).

Beside the explicit $\mathcal{N}=4$ Poincar\'{e} supersymmetry directly
yielded by the considered $\mathcal{N}=4$ superfield formalism, one should
also take into account the additional, implicit $\mathcal{N}=4$
supersymmetry (completing the explicit one to $\mathcal{N}=8$). It is easy
to check that the transformation properties of the above defined $\mathcal{N}%
=4$ superfields read
\begin{equation}
\delta ^{\ast }\hat{\mathcal{V}}_{a\alpha }=\eta _{i\alpha }D_{a}^{i}\hat{%
\mathcal{U}}\;,\quad \delta ^{\ast }\hat{\mathcal{U}}={\frac{1}{2}}\eta
_{i\alpha }D^{ia}\hat{\mathcal{V}}_{a}^{\alpha }\;.  \label{5n4transfa}
\end{equation}

After recalling such facts about the $(\mathbf{5,8,3})$ vector multiplet and
considering the definition (\ref{X}), it is now rather easy to get the
formulations of the new nonlinear $\mathbf{(4,8,4)}$ supermultiplet in terms
of $\mathcal{N}=4$ superfields. Indeed, one just needs to introduce the new $%
\mathcal{N}=4$ superfields
\begin{equation}
\mathcal{L}_{\alpha a}\equiv \frac{{\hat{\mathcal{V}}}_{\alpha a}}{\hat{%
\mathcal{U}}},\;\mathcal{W}^{ia}\equiv \frac{D^{ia}{\hat{\mathcal{U}}}}{{%
\hat{\mathcal{U}}}}.  \label{nN4}
\end{equation}
By rewriting the basic $\mathcal{N}=4$ constraints (\ref{5constra}) in terms
of such $\mathcal{N}=4$ superfields, one obtains
\begin{gather}
D^{i(a}\mathcal{L}^{b)\alpha }+\mathcal{L}^{\alpha (a}\mathcal{W}^{b)i}=0,
\label{new4con1} \\
\notag \\
D^{i(a}\mathcal{W}^{b)i}+\mathcal{W}^{i(a}\mathcal{W}^{jb)}=0\;.
\label{new4con2}
\end{gather}
It is then immediate to recognize that the constraints (\ref{new4con1})
describe a nonlinear version of the $\mathbf{(4,4,0)}$ multiplet, while the
constraints (\ref{new4con2}) define a nonlinear version of the $\mathbf{%
(0,4,4)}$ supermultiplet. The transformations of $\mathcal{L}_{\alpha a}$
and $\mathcal{W}^{ia}$ under the implicit $\mathcal{N}=4$ supersymmetry may
be easily found by recalling their definition (\ref{nN4}) and using Eq. (\ref
{5n4transfa})
\begin{eqnarray}  \label{n4im}
\delta ^{\ast }\mathcal{L}_{\alpha a} &=&\eta _{i\alpha }\mathcal{W}_{a}^{i}-%
\frac{1}{2}\eta _{j\beta }\mathcal{L}_{\alpha a}\left( D^{jc}\mathcal{L}%
_{c}^{\beta }+\mathcal{L}_{c}^{\beta }\mathcal{W}^{jc}\right) ; \\
\delta ^{\ast }\mathcal{W}^{ia} &=&-\frac{1}{2}\eta _{j\alpha }D^{ia}\left(
D^{jb}\mathcal{L}_{b}^{\alpha }+\mathcal{L}_{b}^{\alpha }\mathcal{W}%
^{jb}\right) .  \notag
\end{eqnarray}
Thus we see that our nonlinear $\mathbf{(4,8,4)}$ supermultiplet is
constructed from two $\mathcal{N}=4$ nonlinear supermultiplets, both of which were
never considered before. It also becomes clear what is the role of the
dilaton in the ``linearization'' of our supermultiplet. Indeed,
representing the fermionic superfield $\mathcal{W}^{ia}$ as in (\ref{nN4})
one may easily ``linearize'' both constraints (\ref{new4con1}) and (\ref
{new4con2}), while keeping the fermionic superfield $\mathcal{W}^{ia}$
independent there is no way to have a linear supermultiplet.

\setcounter{equation}0

\section{Analysis of the bosonic sector of the action}

As usual, for constructing the most general superfield action for nonlinear $%
\mathbf{(4,8,4)}$ supermultiplet one should start from the general Ansatz
for the $\mathcal{N}=4$ superfield Lagrangian and impose its invariance with
respect to implicit $\mathcal{N}=4$ supersymmetry (\ref{n4im}). This is not so
easy because among $\mathcal{N}=4$ superfields spanning the $\mathbf{(4,8,4)}$
supermultiplet there are bosonic $\mathcal{L}_{\alpha a}$ and fermionic $%
\mathcal{W}^{ia}$ superfields.

The starting point for the dimensional reduction procedures outlined in
Subsect. 3.1 is the most general sigma-model type action for the $\mathbf{%
(5,8,3)}$ supermultiplet written in the terms of the $\mathcal{N}=4$
superfields defined in (\ref{n4comp}) \cite{bikl1}
\begin{equation}
S=\kappa \int dtd^{4}\theta \mathcal{L}(\hat{\mathcal{V}}_{a\alpha },\hat{%
\mathcal{U}})\;,  \label{raction}
\end{equation}
with the additional constraint that the Lagrangian $\mathcal{L}$ be a harmonic
function
\begin{equation}
\frac{\partial ^{2}\mathcal{L}}{\partial \hat{\mathcal{V}}^{a\alpha
}\partial \hat{\mathcal{V}}_{a\alpha }}+2\frac{\partial ^{2}\mathcal{L}}{%
\partial \hat{\mathcal{U}}^{2}}=0\;.  \label{rinv}
\end{equation}
Performing the $\theta $-integration in (\ref{raction}) and disregarding all
fermionic terms, one obtains the bosonic action
\begin{equation}
S_{B}=-6\kappa \int dtg(v_{a\alpha },u)\left[ {\dot{u}}{}^{2}+2{\dot{v}}%
{}^{a\alpha }{\dot{v}}{}_{a\alpha }-\frac{1}{8}C^{ij}C_{ij}\right] \;,
\label{ractionb}
\end{equation}
where
\begin{equation}
\left. v_{a\alpha }\equiv \hat{\mathcal{V}}_{a\alpha }\right| _{\theta
=0},\;\left. u\equiv \hat{\mathcal{U}}\right| _{\theta =0},\;\left.
C^{ij}\equiv D^{(ia}D_{a}^{j)}\hat{\mathcal{U}}\right| _{\theta =0}
\label{rdef}
\end{equation}
and the metric $g(v_{a\alpha },u)$ of the 5-dim. physical bosonic manifold
is defined as
\begin{equation}
\left. g(v_{a\alpha },u)\equiv \frac{\partial ^{2}\mathcal{L}}{\partial \hat{%
\mathcal{V}}^{a\alpha }\partial \hat{\mathcal{V}}_{a\alpha }}\right|
_{\theta =0}  \label{metric}
\end{equation}
and obeys the constraints
\begin{equation}
\frac{\partial ^{2}g}{\partial v^{a\alpha
}\partial v_{a\alpha }}+2\frac{\partial^{2}g }{%
\partial u^{2}}=0\;.  \label{harm}
\end{equation}
One may wonder whether we can learn something from all this for the cases
of $\mathbf{(4,8,4)}$ supermultiplets, keeping in mind the existence of the
reductions from $\mathbf{(5,8,3)}$ to $\mathbf{(4,8,4)}$. Now we are going
to demonstrate that starting from (\ref{ractionb}) we are able to construct
the most general sigma model actions for $\mathbf{(4,8,4)}$ supermultiplets
together with a particular potential term, in full analogy with $\mathcal{N}%
=4$ supersymmetric cases \cite{root}. For the sake of simplicity, we will
consider only bosonic sectors. The fermionic terms can be easily restored,
if needed.

\subsection{Reduction to the linear $\mathbf{(4,8,4)}$ supermultiplet}

Such a reduction corresponds to constraining the metric $g(v_{a\alpha },u)$
to be independent on $u$
\begin{equation}
g(v_{a\alpha },u)=g_{1}(v_{a\alpha })\;.  \label{r1}
\end{equation}
This functional restriction, when inserted in Eq. (\ref{rinv}), allows one
to write the Lagrangian density $\mathcal{L}(\hat{\mathcal{V}}_{a\alpha },%
\hat{\mathcal{U}})$ in (\ref{raction}) as
\begin{equation}
\mathcal{L}(\hat{\mathcal{V}}_{a\alpha },\hat{\mathcal{U}})=f_{1}(\hat{%
\mathcal{V}})+f_{2}(\hat{\mathcal{V}})\hat{\mathcal{U}}+f_{3}(\hat{\mathcal{V%
}})\hat{\mathcal{U}}{}^{2},
\end{equation}
with the additional constraints
\begin{equation}
\left. \frac{\partial ^{2}{f_{1}}}{\partial \hat{\mathcal{V}}^{a\alpha
}\partial \hat{\mathcal{V}}_{a\alpha }}\right| _{\theta =0}=g_{1},\;\left.
\frac{\partial ^{2}{f_{2}}}{\partial \hat{\mathcal{V}}^{a\alpha }\partial
\hat{\mathcal{V}}_{a\alpha }}\right| _{\theta =0}=0,\;\left. f_{3}\right|
_{\theta =0}=-\frac{1}{4}g_{1},\;\frac{\partial ^{2}{g_{1}}}{\partial
v^{a\alpha }\partial v_{a\alpha }}=0.  \label{r2}
\end{equation}
Thus, in order to perform the reduction to the linear $\mathbf{(4,8,4)}$
supermultiplet, the metric $g_{1}(v_{a\alpha })$ must obey the 4-dimensional Laplace
equation.

Next, we follow the same procedure exploited in the $\mathcal{N}=4$ case in
\cite{root}. We replace $\dot{u}$ by a new auxiliary field $B$ in the action
(\ref{ractionb}) and add the simplest Fayet-Iliopoulos (FI) term (linear in $%
B$)
\begin{equation}
S_{1}=-6\kappa \int dtg_{1}(v_{a\alpha })\left[ B^{2}+2{\dot{v}}{}^{a\alpha }%
{\dot{v}}{}_{a\alpha }-\frac{1}{8}C^{ij}C_{ij}\right] -6\kappa \int dtmB.
\label{ractionb1}
\end{equation}
Eliminating the auxiliary fields in (\ref{ractionb1}) by their equations of
motion, one obtains the following action for physical bosonic components:
\begin{equation}
S_{1}=-12\kappa \int dt\left[ g_{1}{\dot{v}}{}^{a\alpha }{\dot{v}}%
{}_{a\alpha }-\frac{m^{2}}{8g_{1}}\right] .  \label{ractionb2}
\end{equation}
The action (\ref{ractionb2}) corresponds to the general action for the (4,8,4)
linear supermultiplet \cite{anton} with the specific potential term.

\subsection{Reduction to the nonlinear $\mathbf{(4,8,4)}$ supermultiplet}

In order to perform the reduction to the nonlinear supermultiplet, it is
convenient to introduce the new variables
\begin{equation}
l^{a\alpha }\equiv \frac{v^{a\alpha }}{u},\quad y\equiv \frac{\dot{u}}{u}%
,\quad {\hat{C}}{}^{ij}\equiv \frac{C^{ij}}{u}.
\end{equation}
By substituting such definitions in the action (\ref{ractionb}), one gets
\begin{equation}
S_{2}=-6\kappa \int dtg(v_{a\alpha },u)u^{2}\left[ (1+2l^{2})y^{2}+2{\dot{l}}%
{}^{a\alpha }{\dot{l}}{}_{a\alpha }+4yl^{a\alpha }{\dot{l}}{}_{a\alpha }-%
\frac{1}{8}{\hat{C}}{}{}^{ij}{\hat{C}}{}_{ij}\right] \;.  \label{ractionb3}
\end{equation}
It is easy to conclude that the action (\ref{ractionb3}) will correspond to
the nonlinear $\mathbf{(4,8,4)}$ supermultiplet iff
\begin{equation}
g(v_{a\alpha },u)u^{2}=g_{2}(l^{a\alpha }).  \label{rcond1}
\end{equation}
Eq. (\ref{rinv}) implies the metric $g_{2}\left( l\right) $ to satisfy the
following differential equation:
\begin{equation}
\frac{\partial ^{2}}{\partial l^{a\alpha }\partial l_{a\alpha }}g_{2}\left(
l\right) +2l^{a\alpha }l^{b\beta }\frac{\partial ^{2}}{\partial l^{a\alpha
}\partial l^{b\beta }}g_{2}\left( l\right) +12l^{a\alpha }\frac{\partial }{%
\partial l^{a\alpha }}g_{2}\left( l\right) +12g_{2}\left( l\right) =0.
\label{eqqq}
\end{equation}
When the condition (\ref{rcond1}) is fulfilled, one can introduce the
simplest Fayet-Iliopoulos term (linear in $y$)
\begin{equation}
-6\kappa \int dt \; m \;y
\end{equation}
and eliminate the auxiliary field $y$ by its equation of motion, obtaining
the following bosonic action:
\begin{equation}
S_{2}=-12\kappa \int dt\left\{ g_{2}\left[ {\dot{l}}{}^{a\alpha }{\dot{l}}%
_{a\alpha }-2\frac{\left( l^{a\alpha }{\dot{l}}_{a\alpha }\right) ^{2}}{%
1+2l^{2}}\right] -\frac{m^{2}}{8}\frac{1}{g_{2}(1+2l^{2})}\right\}
\label{ractionb4}
\end{equation}
Moreover, by defining the new fields
\begin{equation}
z^{a\alpha }\equiv \frac{\sqrt{2}}{1+\sqrt{1+2l^{2}}}l^{a\alpha },
\label{zzz}
\end{equation}
one can rewrite the action (\ref{ractionb4}) in the following nice form:
\begin{equation}
S_{2}=-24\kappa \int dt\left[ \frac{g_{2}}{(1-z^{2})^{2}}\;{\dot{z}}%
{}^{a\alpha }{\dot{z}}_{a\alpha }-\frac{m^{2}}{16}\frac{(1-z^{2})^{2}}{%
g_{2}(1+z^{2})^{2}}\right]  \label{ractionb5}
\end{equation}
It is interesting to notice that, by performing the change of variable (\ref
{zzz}), the differential equation (\ref{eqqq}) can be rewritten in the
remarkably simple form
\begin{equation}
\frac{\partial ^{2}}{\partial z^{a\alpha }\partial z_{a\alpha }}\left[ \frac{%
1+z^{2}}{(1-z^{2})^{2}}g_{2}\left( z\right) \right] =0,
\end{equation}
which is nothing but the 4-dim. Laplace equation
\begin{equation}  \label{nn}
\frac{\partial ^{2}\mathcal{G}\left( z\right) }{\partial z^{a\alpha
}\partial z_{a\alpha }}=0
\end{equation}
for the redefined metric function
\begin{equation}
\mathcal{G}\left( z\right) \equiv \frac{1+z^{2}}{(1-z^{2})^{2}}g_{2}\left(
z\right) .  \label{redef}
\end{equation}
By inserting the redefinition (\ref{redef}) in the action (\ref{ractionb5}),
one finally gets
\begin{equation}
S_{2}=-24\kappa \int dt\left[ \frac{\mathcal{G}}{1+z^{2}}\;{\dot{z}}%
{}^{a\alpha }{\dot{z}}_{a\alpha }-\frac{m^{2}}{16}\frac{1}{\mathcal{G}\left(
1+z^{2}\right) }\right] .
\end{equation}
Thus, we see that the net effect of using the nonlinear $(\mathbf{4,8,4%
})$ supermultiplet is the deformation of the metric and potential term in
the bosonic sector (together with the deformation of the fermionic terms).

Finally, it is interesting to note that the particular solution of (\ref{nn}%
)
\begin{equation}
\mathcal{G} = \frac{1+z^{2}}{z^2}
\end{equation}
gives rise to the action
\begin{equation}  \label{nnn}
S_{2}=-24\kappa \int dt\left[ \frac{1}{z^{2}}\;{\dot{z}}{}^{a\alpha }{\dot{z}%
}_{a\alpha }-\frac{m^{2}}{16}\frac{z^2}{\left( 1+z^{2}\right)^2 }\right] .
\end{equation}
The metric $\frac{1}{z^2}$ is the solution of the four dimensional Laplace
equation and therefore the sigma-model part of the action (\ref{nnn})
coincides with the action (\ref{ractionb2}) for the linear $(\mathbf{4,8,4})$
supermultiplet with $g_1=\frac{1}{z^2}$. Nevertheless, the potential term in
(\ref{nnn}) is completely different. \setcounter{equation}0

\section{Conclusions}

In this paper we constructed a new nonlinear off-shell $\mathcal{N}{=}8$
supermultiplet with $(\mathbf{4,8,4})$ components content. We showed that
this multiplet can be described in a $\mathcal{N}{=}8$ superfield form as
properly constrained Goldstone superfields associated with suitable cosets
of the nonlinearly realized $\mathcal{N}{=}8,d{=}1$ superconformal group $%
OSp(4^{\star }|4)$. The $\mathcal{N}{=}8$ superfield irreducibility
conditions were derived as a subset of covariant constraints on the Cartan
super one-forms. The superconformal transformation properties of these $%
\mathcal{N}{=}8,d{=}1$ Goldstone superfields were explicitly given,
alongside with the transformation of the coordinates of $\mathcal{N}{=}8,d{=}%
1$ superspace. Although the whole superconformal group $OSp(4^{\star }|4)$ has a
perfect realization on the nonlinear $(\mathbf{4,8,4})$ supermultiplet the
most general action is invariant only under $\mathcal{N}{=}8$ Poincar\'e
supersymmetry.

Apart from the $\mathcal{N}{=}8$ superfield description, we presented also $%
\mathcal{N}{=}4$ superfield formulations of this multiplet. We also
established the relations of this new nonlinear supermultiplet with the linear $(%
\mathbf{5,8,3})$ one. More concretely, there exist reductions from $(\mathbf{%
5,8,3})$ to $(\mathbf{4,8,4})$ linear and nonlinear supermultiplets.
Moreover, these reductions being applied to the action give rise to the most
general sigma-model type action for $(\mathbf{4,8,4})$ supermultiplets with
some sort of potential terms.

The present considerations provide another proof of the statement that the $%
\mathcal{N}{=}4,8$ supermultiplets which do not contain the dilaton among their
components fields are all nonlinear. In this respect, it seems interesting to
analyse the nonlinear supermultiplets related with the other $\mathcal{N}{=}%
8,d=1$ superconformal groups $OSp(8|2),F(4)$ and $SU(1,1|4)$ \cite
{{VP},{FRS}}. The corresponding $\mathcal{R}$-symmetries groups are $%
SO(8),SO(7)$ and $SO(6)$. Therefore one might expect to respectively obtain $%
(\mathbf{7,8,1})$, $(\mathbf{6,8,2})$ and $(\mathbf{5,8,3})$ nonlinear
supermultiplets.

In this paper, when constructing the superfield actions, we preferred to
deal with $\mathcal{N}{=}4, d=1 $ superfields. Thus, only half of the
supersymmetries were manifest. Of course, it would be nice to have a
description with all $\mathcal{N}{=}8$ supersymmetries manifest. This can be
achieved only in harmonic superspace \cite{BI}. \setcounter{equation}0

\section*{Acknowledgements}

Useful discussions with Evgeny Ivanov are gratefully acknowledged.

The work of S.B. has been supported in part by the European Community Human
Potential Program under contract MRTN-CT-2004-005104 ``Constituents,
fundamental forces and symmetries of the Universe''.

The work of S.K. has been supported by RFBR-06-02-16684 and DFG 436 Rus
113/669/0-3 grants.

The work of A.M. has been supported by a Junior Grant of the ``Enrico
Fermi'' Center, Rome, in association with INFN Frascati National
Laboratories.

S.K. would like to thank the INFN--Laboratori Nazionali di Frascati for the
warm hospitality extended to him during the course of this work.


\begin{thebibliography}{99}
\bibitem{ikl1}  E.~Ivanov, S.~Krivonos, O.~Lechtenfeld, Class. Quantum Grav.
21 (2004) 1; \texttt{hep-th/0310299}.

\bibitem{bbkno}  S.~Bellucci, A.~Beylin, S.~Krivonos, A.~Nersessian,
E.~Orazi, Phys. Lett. B616 (2005) 228; \texttt{hep-th/0503244}.

\bibitem{buks}  C.~Burdik, S.~Krivonos, A.~Shcherbakov, Czech. J. Phys. 55
(2005) 1357; \texttt{hep-th/0508165}.

\bibitem{bbks}  S.~Bellucci, A.~Beylin, S.~Krivonos, A.~Shcherbakov, Phys.
Lett. B633 (2006) 382; \texttt{hep-th/0511054}.

\bibitem{ks}  S.~Krivonos, A.~Shcherbakov, Phys. Lett. B637 (2006) 119;
\texttt{hep-th/0602113}.

\bibitem{bks}  S.~Bellucci, S.~Krivonos, A.~Shcherbakov, Phys. Rev. D73
(2006) 085014; texttt{hep-th/0604056}.

\bibitem{GPS}  G.W.~Gibbons, G.~Papadopoulos, K.S.~Stelle, Nucl. Phys.
\textbf{B508} (1997) 623, \texttt{hep-th/9706207}.

\bibitem{N48}  G.~Papadopoulos, Class. Quantum Grav. \textbf{17} (2000)
3715, \texttt{hep-th/0002007};\newline
J.~Michelson, A.~Strominger, Commun. Math. Phys. \textbf{213} (2000) 1,
\texttt{hep-th/9907191}; \newline
J.~Michelson, A.~Strominger, JHEP \textbf{9909} (1999) 005, \texttt{%
hep-th/9908044}; \newline
R.A.~Coles, G.~Papadopoulos, Class. Quantum Grav. \textbf{7} (1990) 427.

\bibitem{N48a}  C.M.~Hull, \textit{``The geometry of supersymmetric quantum
mechanics"}, \texttt{hep-th/9910028}.

\bibitem{N48b}  E.E.~Donets, A.~Pashnev, J. Juan~Rosales, M.M.~Tsulaia,
Phys. Rev. \textbf{D61} (2000) 043512, \texttt{hep-th/9907224}.

\bibitem{bepa}  V.P.~Berezovoj, A.I.~Pashnev, Class. Quantum Grav. \textbf{8}
(1991) 79.

\bibitem{N48c}  S.~Bellucci, E.~Ivanov, A.~Sutulin, Nucl. Phys. \textbf{B722}
(2005) 297, \texttt{hep-th/0504185};\newline
E.A.~Ivanov, A.V.~Smilga, Nucl. Phys. \textbf{B694} (2004) 473, \texttt{%
hep-th/0402041};\newline
E.~Ivanov, O.~Lechtenfeld, JHEP \textbf{0309} (2003) 073, \texttt{%
hep-th/0307111}.

\bibitem{ismi}  E.A.~Ivanov, A.V.~Smilga, Phys. Lett. \textbf{B257} (1991)
79.

\bibitem{leva2}  E.~Ivanov, S.~Krivonos, V.~Leviant, J. Phys. A: Math. Gen.
\textbf{22} (1989) 4201.

\bibitem{N48d}  S.~Bellucci, A.~Nersessian, Phys. Rev. \textbf{D64} (2001)
021702(R), \texttt{hep-th/0101065};\newline
S.~Bellucci and A.~Nersessian, Nucl. Phys. Proc. Suppl. \textbf{102} (2001)
227, \texttt{hep-th/0103005}.

\bibitem{VP}  A.~Van~Proeyen, \textit{Tools for supersymmetry}, \texttt{%
hep-th/9910030}.

\bibitem{FRS}  L.~Frappat, P.~Sorba, A.~Sciarrino, \textit{Dictionary on Lie
superalgebras}, \texttt{hep-th/9607161}.

\bibitem{ikl2}  E.~Ivanov, S.~Krivonos, O.~Lechtenfeld, JHEP 0303 (2003)
014, \texttt{hep-th/0212303}.

\bibitem{bikl1}  S.~Bellucci, E.~Ivanov, S.~Krivonos, O.~Lechtenfeld, Nucl.
Phys. B684 (2004) 321; \texttt{hep-th/0312322}.

\bibitem{bikl2}  S.~Bellucci, E.~Ivanov, S.~Krivonos, O.~Lechtenfeld, Nucl.
Phys. B699 (2004) 226; \texttt{hep-th/0406015}.

\bibitem{286}  S.~Bellucci, S.~Krivonos, A.~Nersessian, Phys.Lett. B605
(2005) 181; \texttt{hep-th/0410029};\newline
S.~Bellucci, S.~Krivonos, A.~Shcherbakov, Phys. Lett. B612 (2005) 283;
\texttt{hep-th/0502245}.

\bibitem{IH}  E.A.~ Ivanov, V.I.~Ogievetsky, Teor. Mat. Fiz. 25 (1975) 164.

\bibitem{root}  S.~Bellucci, S.~Krivonos, A.~Marrani, E.~Orazi, Phys. Rev.
D73 (2006) 025011; \texttt{hep-th/0511249}.

\bibitem{DE}  D.-E.~Diaconescu, R.~Entin, Phys. Rev. D 56 (1997) 8045-8052;
\texttt{hep-th/9706059}.

\bibitem{anton}  S.~Bellucci, S.~Krivonos, A.~Sutulin, Phys. Lett. B605
(2005) 406; \texttt{hep-th/0410276},\newline
S.~Bellucci, E.~Ivanov, A.~Sutulin, Nucl. Phys. B722 (2005) 297; \texttt{%
hep-th/0504185}.

\bibitem{BI}  S.~Bellucci, E.~Ivanov, \textit{in preparation}.
\end{thebibliography}
\end{document}